%% file: bib.tex
\documentclass[manuscript,screen]{acmart}

\AtBeginDocument{%
  \providecommand\BibTeX{{%
    Bib\TeX}}}

\setcopyright{acmlicensed}
\copyrightyear{2018}
\acmYear{2018}
\acmDOI{XXXXXXX.XXXXXXX}

\acmJournal{JACM}
\acmVolume{37}
\acmNumber{4}
\acmArticle{111}
\acmMonth{8}
\usepackage{nicefrac}
\usepackage{siunitx}
\usepackage{array,framed}
\usepackage{pdfpages}
\usepackage{booktabs}
\usepackage{
  color,
  float,
  epsfig,
  wrapfig,
  graphics,
  graphicx,
  subcaption
}
\usepackage{hyperref}
\usepackage{setspace}
\usepackage{amsfonts}
\usepackage{latexsym,fancyhdr}
\usepackage{enumerate}
\usepackage[linesnumbered,ruled,longend,noend]{algorithm2e}
\usepackage{algpseudocode}
\usepackage{graphicx}
\usepackage{xparse} 
\usepackage{xspace}
\usepackage{amsmath} 
\usepackage{multirow}
\usepackage{csvsimple}
\usepackage{balance}
\usepackage{bbding}
\usepackage{flushend}
\usepackage{pifont}
\usepackage{threeparttable}
\usepackage{longtable}
\usepackage{wrapfig}
\usepackage{shortcuts}
\usepackage{adjustbox}
\usepackage{booktabs} 
\usepackage{soul}
\usepackage{todonotes}
\usepackage{enumitem}
\usepackage{array}
\usepackage{color}
\usepackage{mdframed}
\usepackage{hyperref}
\def\BibTeX{{\rm B\kern-.05em{\sc i\kern-.025em b}\kern-.08em
    T\kern-.1667em\lower.7ex\hbox{E}\kern-.125emX}}
\begin{document}

\title{\tool{}: Semantic Preserving Object Layout Reconstruction for Image Captioning System Testing}

\author{Yi Liu}
\affiliation{%
  \institution{Huazhong University of Science and Technology}
  \country{China}}
\affiliation{%
  \institution{Nanyang Technological University}
  \country{Singapore}}
\authornote{These authors contributed equally to this research.}

\author{Guanyu Wang}
\affiliation{%
  \institution{Huazhong University of Science and Technology}
  \country{China}}
\authornotemark[1]

\author{Xinyi Zheng}
\affiliation{%
  \institution{Huazhong University of Science and Technology}
  \country{China}}

\author{Gelei Deng}
\affiliation{%
  \institution{Nanyang Technological University}
  \country{Singapore}}

\author{Kailong Wang}
\affiliation{%
  \institution{Huazhong University of Science and Technology}
  \country{China}}
\authornote{Corresponding author.}

\author{Yang Liu}
\affiliation{%
  \institution{Nanyang Technological University}
  \country{Singapore}}

\author{Haoyu Wang}
\affiliation{%
  \institution{Huazhong University of Science and Technology}
  \country{China}}

\input{Chapters/abstract}

\maketitle

\input{Chapters/introduction}
\input{Chapters/background}
\input{Chapters/motivation}
\input{Chapters/methodology}
\input{Chapters/evaluation}
\input{Chapters/discussion}

\input{Chapters/related-work}
\input{Chapters/conclusion}
\clearpage
\bibliographystyle{ACM-Reference-Format}
\bibliography{bib}

\end{document}

%% file: Chapters/abstract.tex
\begin{abstract}
Image captioning~(IC) systems, including Microsoft Azure Cognitive Service, are commonly utilized to convert image content into descriptive natural language. However, inaccuracies in caption generation can lead to serious misinterpretations. Advanced testing techniques such as \metaic{} and \rome{} have been developed to mitigate these issues, yet they encounter notable challenges. Firstly, these strategies demand intensive labor, relying on detailed manual annotations like bounding box data of objects to create test cases. Secondly, the realism of the generated images is compromised, with \metaic{} adding unrelated objects and \rome{} failing to remove objects effectively. 
Finally, the capability to generate diversified test suites is restricted. \metaic{} is limited to only inserting specific objects to prevent overlap, whereas \rome{} can generate only $3^n - 2^n$ variations of test cases from an original seed image containing $n$ objects.


In this study, we present \tool{}, a novel automated tool designed for semantic preserving object layout reconstruction in image captioning system testing. \tool{} is based on the insight that modifying the arrangement of objects within an image does not alter its inherent semantics. We utilize four semantic-preserving transformation techniques—translation, rotation, mirroring, and scaling—to modify object layouts autonomously, eliminating the need for manual annotation. This approach enables the creation of realistic and varied test suites for IC system testing. Our extensive testing demonstrates that more than 75\% of survey respondents find the images produced by \tool{} more realistic compared to those generated by SOTA methods. Additionally, \tool{} exhibits outstanding performance in identifying caption errors, detecting 31,544 incorrect captions across seven IC systems with an average precision of 91.62\%. This significantly outperforms other methods, which only achieve 85.65\% accuracy on average and identify 17,160 incorrect captions. Notably, \tool{} exposes 6,236 unique issues within Microsoft Azure Cognitive Service, highlighting its effectiveness against one of the most advanced IC systems available.


\end{abstract}

%% file: Chapters/introduction.tex
\section{Introduction}
\label{sec:intro}

\textbf{I}mage \textbf{C}aptioning (\textbf{IC}) systems, which convert images into text, are widely used in various areas such as aiding visually impaired individuals, auto-generating social media captions, improving image search functionalities, and bolstering security surveillance measures. Given their broad application and impact, it is essential to rigorously test these systems to ensure accuracy and reliability, and to prevent errors like misclassifications or omissions in captions. This not only enhances system performance but also builds trust in technology that is becoming integral in our digital interactions.

\begin{figure*}[t]
\centering
\includegraphics[width=\textwidth]{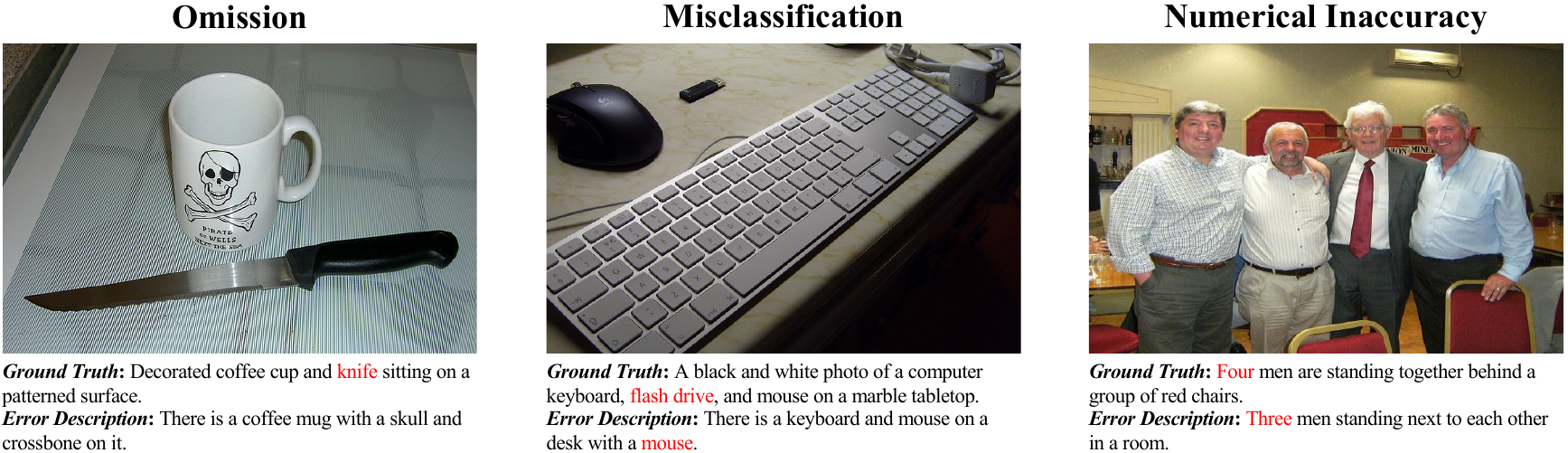}
\caption{Examples of three types of common errors.}
\label{fig:err}
\end{figure*}

Unfortunately, akin to other deep learning-based technologies, even the most sophisticated IC systems are not immune to generating incorrect captions~\cite{metaic, rome}. As exemplified in Figure~\ref{fig:err}, such errors can manifest in various forms. For instance, 
it only describes the cup and its design, overlooking another point that should be noted (\textbf{Omission Error}).
In another case, 
a flash drive is misdescribed as a mouse, indicating a potential \textbf{Misclassification Error}.
Furthermore, a gathering of four men is inaccurately captioned as three, a clear \textbf{Numerical Inaccuracy}. These instances highlight the critical need for ongoing refinement and thorough testing of these AI-driven IC systems to enhance their precision and reliability.

Several approaches have been developed for testing image captioning systems, with techniques like \metaic{}~\cite{metaic} and \rome{}~\cite{rome} standing out. \metaic{} operates by inserting objects into background images, creating synthesized images with varying degrees of overlap. This approach then feeds image pairs into the captioning system to detect potential errors through metamorphic relations. On the other hand, \rome{} takes a different approach. It systematically removes objects from images and uses advanced image inpainting to fill the gaps. As a result, \rome{} generates natural-looking images that are used to test the captioning system's ability to accurately describe images with altered content. 

Existing testing techniques for image captioning systems, including \metaic{} and \rome{}, face three significant limitations. Firstly, there is a heavy reliance on manual efforts (\textbf{{Challenge \#1}}): both \metaic{} and \rome{} require extensive human input to annotate object categorization and positions, making the process time-intensive and costly. Secondly, the issue of image realism (\textbf{{Challenge \#2}}) arises. Despite claims of generating natural-looking images, both techniques struggle in this area. \metaic{}, for instance, might insert irrelevant objects, like placing an elephant on a table, while \rome{} might incorrectly replace a car on the road with a dustbin during its object melting process, resulting in unrealistic images. Lastly, diversity (\textbf{{Challenge \#3}}) is a concern. Given an annotated dataset, these methods can generate only a limited number of test images. For example, \rome{} selects images with a number of objects constrained by a threshold, and the melting process can only produce a certain number of test cases, which fails to cover a broad range of existing image-caption pairs for comprehensive testing.

\noindent\textbf{Our Solution.} Drawing inspiration from the translational invariance property of images~\cite{kayhan2020translation}, we have formulated four metamorphic relations to modify the layout of objects in an image. This approach leads to the generation of new images for testing IC systems. We introduce \tool{}, a novel semantic preserving 
Object Layout Re-constructor for IC systems that employs Metamorphic Testing (MT)~\cite{chen2020metamorphic} based on a series of semantic preserving metamorphic relations, effectively overcoming the challenges mentioned earlier. Specifically, to address the heavy reliance on manual annotations for object semantics, such as positioning (\textbf{Challenge \#1}), \tool{} implements semantic segmentation on image-caption pairs. This process automatically extracts objects and their corresponding masks from images, thereby resolving the first challenge.

Accurate masks are crucial for generating realistic images. To ensure the generated images are realistic (\textbf{Challenge \#2}), we implement four strategies: (1) We recursively melt down objects to obtain accurate masks using inpainting techniques. (2) In cases where objects overlap, leading to incomplete masks from semantic segmentation, we use inpainting to complete these masks. (3) After inpainting, we retain the original background in our subsequent generated images, minimizing perturbations to the original image's semantics. With these refined masks, we apply metamorphic transformations to create new mask layouts and use mask-to-image translation techniques to produce new images for testing in IC systems.

Regarding comprehensiveness (\textbf{Challenge \#3}), our defined metamorphic relations, such as translation, rotation, scaling and mirroring, offer a continuous space for altering object layouts in an image. Thus, starting from a limited number of seed images, we can generate a wide variety of different image layouts for comprehensive testing of IC systems.

\noindent\textbf{Results.}
In this work, we conducted comprehensive evaluations on seven leading IC systems, including Microsoft's Azure~\cite{azure}. With 200 images as seeds, \tool generated 10,000 test cases, while other baselines produced a maximum of only 2,906 without any limitations. Note that this boundary is set considering the difficulty of manual review, and it is not the upper limit of \tool{}'s generation capability.  The survey results show that over 75\% of participants favor \tool{} for producing more realistic images compared to existing methods like \rome{}. Additionally, \tool{} outperforms other SOTA techniques in error detection precision, showing a 
5.97\% increase in error detection across various IC systems under test. An ablation study was conducted to assess the effectiveness of the four metamorphic relations in identifying errors. \tool{} identified a total of 31,544 errors in seven IC systems, with 6,236 errors detected specifically in Azure.

\noindent\textbf{Contribution.}
The key contributions of this work are summarized as follows:

\begin{itemize}[itemsep=0pt,parsep=0pt,topsep=0pt,partopsep=0pt,left=0pt]

\item\textit{New Testing Methodology.}  We introduce object layout editing as a novel and generic methodology for testing image captioning software.

\item\textit{Artifact Availability.} \tool{} is implemented using LaMa~\cite{suvorov2021resolution} for image inpainting, Stanza’s~\cite{qi2020stanza} POS Tagging for object name identification, and PITI~\cite{wang2022pretraining} for mask-to-image translation. We have made \tool{}'s implementation and its generated datasets available on our website~\cite{ourtool} to facilitate future research. 

\item\textit{Comprehensive Evaluation.} A comprehensive evaluation of \tool{} have been conducted, showcasing its effectiveness. The empirical results clearly demonstrate that \tool{} not only surpasses the precision of current state-of-the-art methods, with a remarkable range from 80.93\% to 95.15\%, but it also excels in creating test cases that are statistically validated to be more natural and realistic by our user study.

\end{itemize}

%% file: Chapters/background.tex
\section{Background} 
\label{sec:background}


\subsection{Image Captioning}
Image captioning can be formally defined as a computational task where the goal is to generate a textual description \( D \) for a given image \( I \). This process can be represented by the function: \( D = f(I) \). Here, \( f \) represents the image captioning algorithm, which maps the input image \( I \) to a descriptive caption \( D \). This function encompasses various sub-processes such as feature extraction, object recognition, and natural language processing. The effectiveness of \( f \) depends on its ability to accurately interpret the visual content and context within \( I \) and translate them into a meaningful and coherent description \( D \). The challenge lies in designing \( f \) to handle diverse visual scenes and generate captions that are both accurate and contextually relevant.

Deep learning has become a widely accepted approach for solving the complexities of image captioning, showcasing significant advancements in this field. Such as convolutional neural networks (CNNs)~\cite{lecun1998gradient} are used for extracting visual features from images, and recurrent neural networks (RNNs) or Long Short-Term Memory (LSTM)~\cite{6795963} networks are employed for generating coherent textual descriptions based on these features. These technologies have been instrumental in developing systems like Google's automated image captioning in Google Photos~\cite{sharma2018conceptual}. These models efficiently recognize and describe the content of images. Similarly, Facebook's automatic alternative text generation for images uses deep learning to enhance accessibility for visually impaired users~\cite{metaAI}. 

In this paper, we explore the testing of advanced multimodal pretraining IC systems. We evaluate services and models like Azure AI Vision API~\cite{azure} from Microsoft, which analyzes images and generates captions, and GIT~\cite{wang2022git}, a generative image-to-text transformer. BLIP~\cite{li2022blip} and its successor BLIP2~\cite{li2023blip2} represent vision-language pretraining frameworks utilizing web data and large language models. ViT-GPT2~\cite{kumar2022illustrated} merges vision transformer (ViT) with GPT2~\cite{radford2019language} for vision-language tasks. OFA~\cite{wang2022ofa} integrates diverse vision and language tasks in a unified learning framework. VinVL~\cite{zhang2021vinvl} focuses on enhancing visual representations in visual-language models. Our work explores the robustness of these state-of-the-art models and applications in the image-to-text domain.


\vspace{-1mm}
\subsection{Testing for IC Systems}
Several pioneers~\cite{metaic, rome} in the field have developed testing techniques for IC systems using metamorphic testing. Metamorphic testing~\cite{chen2020metamorphic} is a methodology that addresses the challenge of testing software in situations where there is no clear oracle to determine the correct output. It relies on metamorphic relations, which are necessary properties of the target function that remain invariant under certain transformations of the input. In the context of IC systems, these relations help in identifying discrepancies in the system's output by analyzing how changes in the input image affect the generated captions. 

Existing research has leveraged metamorphic testing for evaluating IC systems. Specifically, \metaic{}~\cite{metaic} employs a unique strategy where it manipulates input images by inserting objects, and then observes the changes in the output captions to assess the system's accuracy. However, a limitation of \metaic{} is its tendency to create unnatural or contextually inappropriate images, which might not accurately represent real-world scenarios. To overcome the limitations of \metaic{}, \rome{}~\cite{rome} was proposed. \rome{} adopts a different approach by removing objects from images and employing advanced inpainting techniques to maintain natural image context. Despite this improvement, \rome{} has its own limitations, particularly in the realm of generating a diverse range of test scenarios, as its methodology can be constrained by the types of objects and their interactions within the given images.

\vspace{-1mm}
\subsection{Image-to-Image Translation}

Image-to-Image translation~\cite{Isola_2017_CVPR} refers to a category of issues in the fields of graphics and computer vision. Its goal is to create images in the target domain that follow the input's semantics accurately. This concept gained prominence with the introduction of Generative Adversarial Networks (GANs) by ~\cite{goodfellow2014generative}. Following developments like Pix2Pix~\cite{Isola_2017_CVPR} and CycleGAN~\cite{zhu2017unpaired} have shown that conditional GANs can be effectively used to convert images from one domain to another, such as turning sketches into photographs or converting black and white images into color. 

More recently, diffusion models have demonstrated exceptional ability when trained using large-scale text-image pairs. Instead of directly generating images, diffusion models gradually transform a distribution of random noise into the distribution of real images over a sequence of steps. The key advantage of diffusion models is their ability to generate more diverse and high-quality images compared to earlier techniques. ~\cite{saharia2022palette} has demonstrated the use of diffusion models for image-to-image translation, particularly in data-rich scenarios like image colorization. ~\cite{wang2022pretraining} illustrates how a well-pretrained diffusion model can act as a universal generative prior, facilitating various synthesis tasks based on these advancements. This progress forms a solid base for us for generating high-quality test cases in image synthesis.

%% file: Chapters/motivation.tex
\section{Motivation}
\label{sec:motivation}

\begin{figure}[t]
  \centering
  \includegraphics[width=\linewidth]{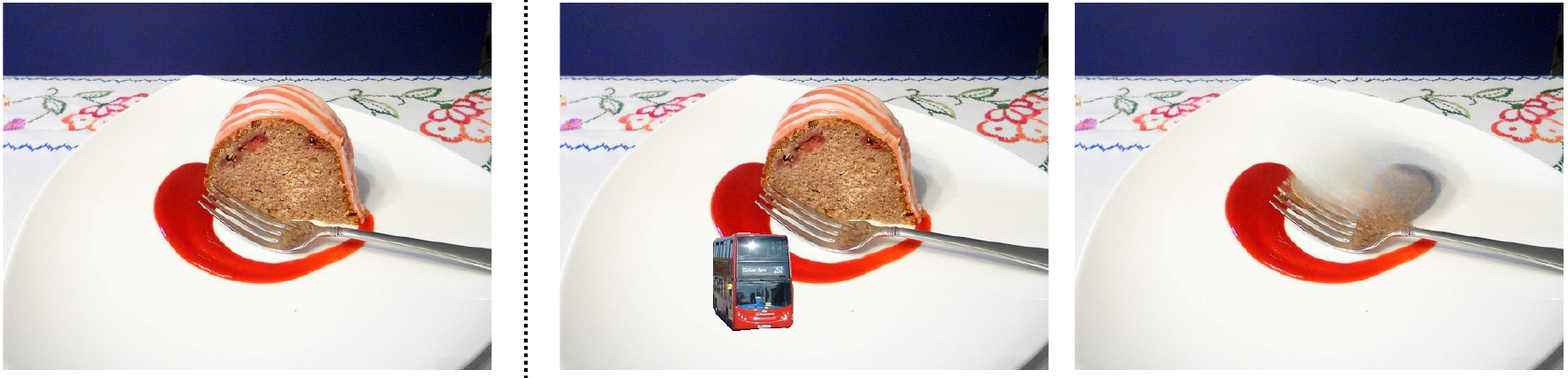}
  \caption{Unrealistic examples generated by existing metamorphic testing frameworks for IC systems. The left is the seed image. The middle and right come from \metaic and \rome.}
  \label{fig:motivation}
\end{figure}


As stated in \autoref{sec:intro}, the current SOTA IC system testing methods like \metaic and \rome still have limitations. This section will discuss these limitations that motivate our work.


\noindent\textbf{Motivation \#1: Intensified Manual Efforts}. Both \metaic{} and \rome{} demand significant manual effort to create images for testing IC systems. Specifically, \metaic{} necessitates precise object positioning to determine the location for inserting objects. Similarly, \rome{} also requires accurate object location information for its object melting process and subsequent image inpainting.


\noindent\textbf{Motivation \#2: Limited Realism of Generated Images}. Both \metaic{} and \rome{} strive to produce realistic images, but they often fall short. \metaic{}, for instance, might insert irrelevant objects into scenes, such as placing a bus on a plate as depicted in \autoref{fig:motivation}. Similarly, \rome{} can create unnatural images; an example in \autoref{fig:motivation} shows how after melting down a meatloaf, \rome{} leaves a blurred background without adequately inpainting the surroundings.


\noindent\textbf{Motivation \#3: Limited Diversity of Generated Images}. The diversity of generated images is a significant constraint in existing approaches. These methods generate test cases by simply inserting or deleting objects, which limits their generative capacity.

%% file: Chapters/methodology.tex
\section{Methodology And Implementation}
\label{sec:methodology}

\begin{figure*}[t]
\centering
\includegraphics[width=\textwidth]{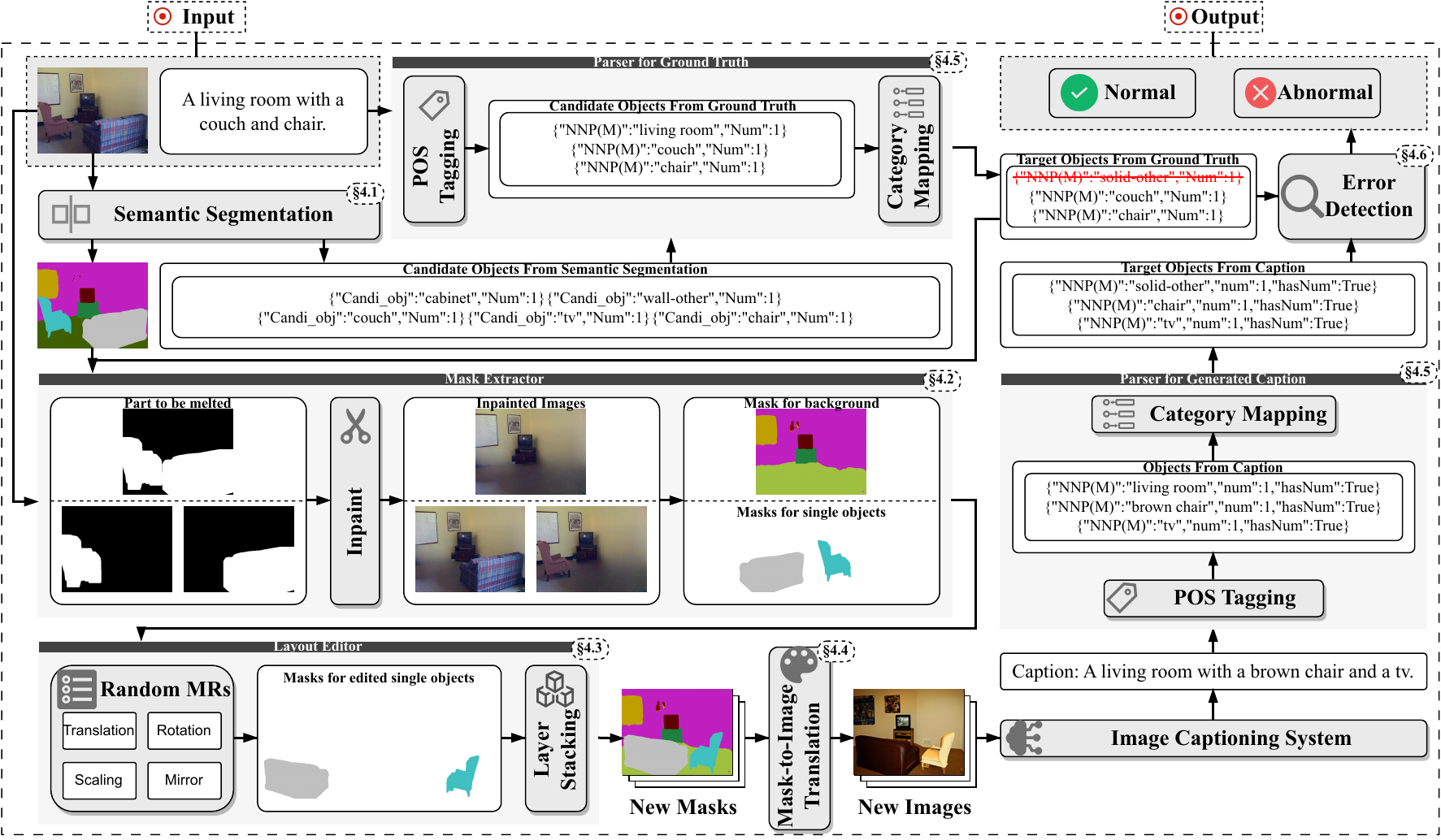}
\caption{\tool overview with two major parts: Image Processing and Text Processing.}
\label{fig:overview}
\end{figure*}

In this section, we present \tool{} and its implementation details. The key concept of \tool{} is driven by the principle of image translation invariant~\cite{kayhan2020translation}, which suggests that altering the layout of an image does not compromise its semantic integrity. \tool{} takes image-caption pairs as inputs and produces outputs that include a series of images transformed by four distinct metamorphic relations. \tool{} highlights image-caption pairs are considered as violations generated by IC system under test. \autoref{fig:overview} illustrates the workflow of \tool{}, detailing each step in this transformative process as following:

\begin{itemize}[itemsep=0pt,parsep=0pt,topsep=0pt,partopsep=0pt,left=0pt]
    \item \textbf{Semantic Segmentation} (\S~\ref{sec:semseg}): For each image input, we implement semantic segmentation to identify the types of objects present and generate their corresponding masks.
    
    \item \textbf{Mask Extractor:} (\S~\ref{sec:splitter}): We utilize a novel inpainting-based extraction algorithm for recursively extracting accurate masks of objects, ensuring precise object delineation.
    
    \item \textbf{Layout Editor:} (\S~\ref{sec:editor}): Utilizing the obtained object masks, we perform transformations based on metamorphic relations to generate a complete mask with a varied layout.
    
    \item \textbf{Mask-to-Image Translation:} (\S~\ref{sec:translator}): With the complete mask in hand, we employ a diffusion model to render a new image that aligns with the modified layout. This image is then processed by the IC system under test.
    
    \item \textbf{Caption Parser}  (\S~\ref{sec:parser}): \tool{} parses the generated captions using Part-Of-Speech (POS) tagging to identify objects and their quantities.

    \item \textbf{Error Detection} (\S~\ref{sec:judgement}) : We then compare these elements with the ground truth. Any discrepancies in object types or count are flagged as violations.
\end{itemize}



\vspace{-1mm}
\subsection{Semantic Segmentation}
\label{sec:semseg}
In \tool{}, since the modification of object layouts in a given image is crucial, this step focuses on extracting position information, specifically the masks, and the associated objects. To achieve this, we initially conduct semantic segmentation on the input images. This process generates their semantic maps, providing detailed information about the spatial distribution (shape and bounding box) and characteristics (object name and its count) of various objects within the image.

Specifically, for segmenting the input images, we utilize OpenSeeD~\cite{zhang2023simple}, a framework designed for Open-Vocabulary Segmentation and Detection. Our segmentation and detection target the categories listed in the MSCOCO~\cite{lin2014microsoft} and COCO-Stuff~\cite{Caesar_2018_CVPR}. 
These categories also align with the classes trained in image translation model we used.
From the segmentation results, we identify and extract the objects, applying a predefined color mapping to their semantic masks. To address the issue of vague descriptions in ground truths, such as ``a group of'', which often leads to challenges in detecting quantity errors, we meticulously count the number of each object type from the segmentation results. Consequently, we obtain unprocessed semantic masks along with candidate sets of target objects, setting the stage for further processing in \tool{}.

\subsection{Mask Extractor}
\label{sec:splitter}

\input{Chapters/algo-splitter}

Building upon the semantic masks obtained in the previous step, as detailed in \autoref{sec:semseg}, we encounter two critical issues before we can modify the layout of the original image:

\noindent\textbf{Problem 1: Selection of Modifiable Objects.} The primary concern is determining which objects in an image can have their layouts modified. With multiple objects present, it is crucial to identify specific objects whose positional changes will not completely alter the image's semantic meaning.

\begin{figure}[t]
    \centering
    \begin{subfigure}{\textwidth}
        \centering
        \includegraphics[width=0.8\textwidth]{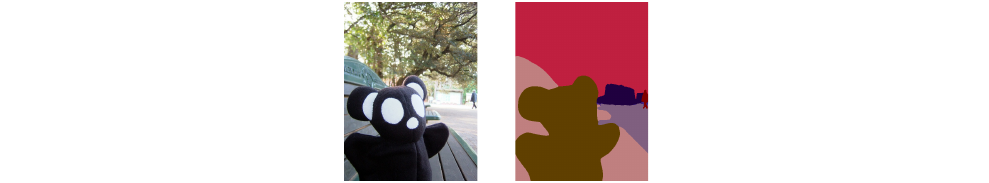}
        \caption{The original image(Left) and its semantic map(Right).}
        \label{fig:ori&mask}
    \end{subfigure}
    \begin{subfigure}{\textwidth}
        \centering
        \includegraphics[width=0.8\textwidth]{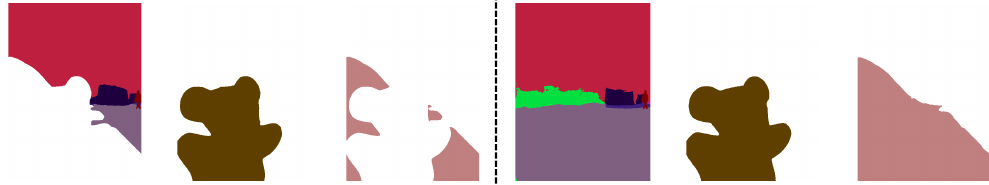}
        \caption{Masks for background and single objects. The left shows the results of direct segmentation, and the right is the results of the segmentation after inpainting.}
        \label{fig:masks}
    \end{subfigure}
    \begin{subfigure}{\textwidth}
        \centering
        \includegraphics[width=0.8\textwidth]{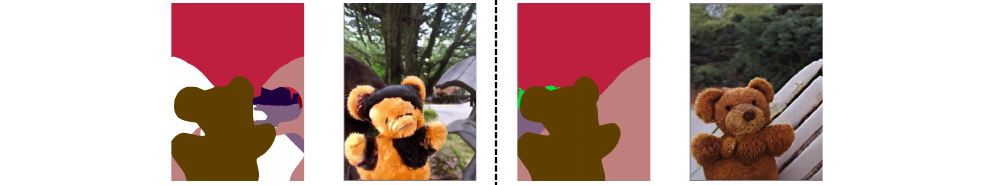}
        \caption{Semantic maps with new layout and images based on them.}
        \label{fig:gen}
    \end{subfigure}
    \caption{The impact of inpainting image on generating new images when extracting masks.}
    \label{fig:issue}
\end{figure}

\noindent\textbf{Problem 2: Acquisition of Accurate Object Masks.} As illustrated in \autoref{fig:issue}, a prevalent challenge emerges when objects in an image overlap or are stacked, as shown in Figure~\ref{fig:issue}(a) where the original image and its semantic map reveal the spatial relations of objects. In scenarios such as this, the semantic segmentation task becomes complex, leading to incomplete or inaccurate masks, exemplified by Figure~\ref{fig:issue}(b). The left side of this subfigure demonstrates the segmentation result directly obtained from the original image, which may include parts of overlapping objects, while the right side shows the improved segmentation after inpainting. These inaccuracies in mask acquisition are crucial to address, as they can significantly deteriorate the quality of the resultant image, demonstrated by Figure~\ref{fig:issue}(c). Here, the semantic maps with the new layout and the corresponding images synthesized based on them indicate that obtaining precise and distinct object masks is essential for generating high-fidelity images, particularly when dealing with object occlusions or complex arrangements.

To tackle \textbf{Problem 1}, we employ a two-fold criterion for selecting objects whose layout can be modified. Firstly, we opt for objects that intersect between those segmented in the image and those listed in the ground truth (We present how to extract objects in ground truth captions in Section~\ref{sec:parser}). This ensures that only relevant and accurately identified objects are considered for layout modification. Additionally, we refrain from applying transformations to the original background, maintaining the original semantic context of the images.

In addressing \textbf{Problem 2}, we recognize the issue of incomplete masks arising from overlapping objects. To mitigate this, we implement an inpainting technique on these incomplete masks. This approach is designed to yield complete and distinct masks for each object, even in scenarios where objects are stacked or closely positioned. This step is crucial for ensuring the integrity and clarity of the transformed images.

After getting the semantic maps of seeds and the relevant feature information of the objects, we edit the layout of the objects by the following step (we present it in Section~\ref{sec:editor}). To facilitate adjusting the spatial position of the objects, we divide the mask map to extract the masks of individual objects.

The process for extracting semantic masks for each individual target object is detailed in \autoref{algo:splitter}. Specifically, we begin by creating an empty image template. Then, we meticulously iterate over each pixel in the original semantic map. If a pixel's color aligns with that of any object except the one currently under consideration, we mark that pixel as white. In contrast, pixels matching the current object's color are marked as black. The white regions represent areas that need to be excised and subsequently repaired in the original image. Before executing the repair process, we expand these white areas and perform a backfill operation tailored to the current object. This is done with the intent to preserve the integrity of the current object while eliminating other objects as effectively as possible. For the inpainting of these initial images, we utilize the LaMa~\cite{suvorov2021resolution}. This is followed by a recursive re-segmentation and extraction of the current object's mask based on its color. The final step involves obtaining the background mask, which is accomplished by similarly removing all target objects from the image.

\subsection{Layout Editor}
\label{sec:editor}
At this stage, our objective is to create new layouts for the masks by applying four distinct metamorphic relations (MRs). These relations are pivotal in transforming the spatial arrangement of objects within the images, while preserving their individual characteristics. We detail each MR as follows:

\noindent \textbf{MR1: Translation.} This relation entails random horizontal and vertical shifts of the target mask, characterized by step lengths $\Delta x$ and $\Delta y$. To ensure that the mask remains within the boundaries of the canvas, $\Delta x$ and $\Delta y$ must adhere to the following constraints:
\begin{align}
    -x_{min} \leq &\Delta x \leq Width - x_{max} \label{eq:limitx} \\
    -y_{min} \leq &\Delta y \leq Height - y_{max} \label{eq:limity}
\end{align}

here, $x_{min}$, $x_{max}$, $y_{min}$, and $y_{max}$ denote the extremities of the current object's coordinates in all directions, while $Width$ and $Height$ are the dimensions of the canvas. These boundary values shift with each step.

\noindent \textbf{MR2: Rotation.} We determine the rotational center $C_{obj}$ of the object as per \autoref{eq:center} and rotate it by a random angle $\theta$, either clockwise or counterclockwise.
\begin{align}
    C_{obj} = (\frac{x_{min} + x_{max}}{2}, \frac{y_{min} + y_{max}}{2}) \label{eq:center}
\end{align}

\noindent \textbf{MR3: Scaling.} In a manner akin to \textbf{Rotation}, the scaling center $C_{obj}$ is identified, following which the object is scaled by a random ratio $\alpha$.

\noindent \textbf{MR4: Mirror.} This operation involves a horizontal mirroring of the object along its vertical axis.

Utilizing the four MRs, we adeptly apply them to the masks to generate novel layouts. As detailed in \autoref{algo:editor}, our approach in each step involves a random selection of an object's mask, followed by the application of a predefined MR for editing. This process is iteratively conducted until the allocated step budget is fully utilized. Upon completion, we amalgamate the masks of all individual objects with the background, thereby creating an entirely new semantic map that reflects these transformations.

\input{Chapters/algo-editor}

\vspace{-2mm}
\subsection{Mask-to-Image Translation}
\label{sec:translator}

This step is dedicated to transforming the modified layouts into new images. We employ a mask-to-image translation technique for this image generation process. The essence of mask-to-image translation lies in its ability to realistically generate images, which necessitates the model’s understanding of both structural and textual information inherent in natural scenes. In \tool{}, PITI~\cite{wang2022pretraining} is selected as our default mask-to-image model, due to it is the SOTA mask-to-image model.

Specifically, the modified mask, along with mask-to-text pairs, are inputted into PITI. PITI then undertakes the translation from masks to images, capitalizing on the provided layout and textual data. Owing to the unique properties of diffusion models utilized in PITI, we are able to generate multiple new and distinct images from a single semantic map. This capability significantly enhances the variety our test cases, offering a broad spectrum of scenarios for thorough evaluation.

\vspace{-2mm}
\subsection{Caption Parser}
\label{sec:parser}

\input{Chapters/algo-parser}

In this section, we shift our focus from image processing to the handling of captions, both from ground truths and those generated by IC systems under test.

In \tool{}, we emphasize verifying the accuracy and quantity of objects within an image. This approach mitigates the impact of image colors on performance, such as the challenges posed by monochrome images and the handling of color in captions. Our focus is on parsing image descriptions to extract and identify detectable objects.

As outlined in \autoref{algo:parser}, the parser's input includes the caption, its source, and information about candidate objects identified via semantic segmentation. The initial step involves Part-Of-Speech (POS) Tagging on the caption to extract sentence segments, employing Stanza~\cite{qi2020stanza} for this purpose. Stanza is a versatile NLP library supporting multiple languages for NLP tasks, developed by the Stanford NLP group. The goals extracted include \textbf{nouns or noun phrases (NNP)} and \textbf{nouns or noun phrases with modifiers (NNPM)}.

\begin{definition}
    \textbf{(NNP(M))} Nouns or noun phrases (with modifiers) are defined as an N-tuple:
    \begin{center}
        NNP(M) = $<(ADJ_i), NN_{i+1}, ..., NN_{k}>$,
    \end{center}
\end{definition}
where $NN_{i+1}$ and $NN_{i+k}$ are consecutive nouns and $ADJ_{i}$ is the modifying adjective preceding $NN_{i+1}$ in a caption.


Next, we determine the count of the current NNP(M) and assess if its quantity description needs error detection in the result comparison, particularly if a definite quantifier precedes it. For captions from ground truths, we process all extracted NNP(M)s and map them to predefined categories using a classifier. Here, we employ a Bart-based~\cite{lewis2019bart} zero-shot sequence classifier for this mapping. The process iteratively refines the range of candidate categories until the final one is determined. If the category falls within those extracted via semantic segmentation, we finalize the information and add it to the list of target objects. Conversely, for captions from tested models and applications, we retain all NNP(M)s and map them directly to predefined categories.

\vspace{-2mm}
\subsection{Error Detection}
\label{sec:judgement}

This subsection outlines our approach for detecting errors in IC systems under test. A key observation guiding our method is that editing the layout of objects does not alter the type and number of objects in an image. Leveraging this insight, we execute error detection based on the following relation:



\vspace{-5mm}
\begin{align}
    S_{gt}[Obj] \subseteq S_{gen}[Obj] \label{eq:judge}
\end{align}
\vspace{-5mm}

where $S_{gt}$ and $S_{gen}$ denote the sets of $Obj$ in the ground truth and generated captions, respectively. $Obj$ is an object information dictionary containing $type$ and $num$ attributes. Specifically, $Obj$ in $S_{gen}$ includes an additional $hasNum$ tag to determine whether to perform a detection for the quantity. 
Considering that different IC systems have different granularity in describing the parts outside the image subject object, there should be an inclusion relationship between $S_{gt}$ and $S_{gen}$. 
This implies that the objects identified in the generated captions should completely encompass those extracted from the ground truths. When there is a specific count associated with an object, a further comparison of quantities is conducted. Any mismatches are noted as errors.

%% file: Chapters/algo-splitter.tex
\SetKwFunction{MapSplit}{MapSplit}
\SetKwFunction{Dilate}{Dilate}
\SetKwFunction{ObjSplit}{ObjSplit}
\SetKwFunction{MaskGen}{MaskGen}
\SetKwProg{Fn}{Function}{:}{}

\begin{algorithm}[t]
\scriptsize
\caption{Mask Extractor}\label{algo:splitter}
\KwData{Seed Image $img$, Semantic Map of the Seed Image $mask$ and Objects to be detected $targets$}
\KwResult{Masks for individual objects $single$}

\Fn{\MapSplit{img, mask, targets}}{
    \tcp{Extract masks for individual objects}
    \For{$cur$ \textbf{in} $ targets$}{
        $\MaskGen{img, mask, targets, "None", cur}$
    }
    \tcp{Extract the mask for background}
    $\MaskGen{img, mask, targets, "Background", "None"}$
}

\Fn{\MaskGen{img, mask, targets, type, cur}}{
    $inpaint\_mask \gets zeros\_like(mask)$\;
    \For{$pixel$ \textbf{in} $mask$}{
        \eIf{$pixel \in targets$ \textbf{and} (($cur \neq None$ \textbf{and} $pixel \notin cur$) \textbf{or} $type$ \textbf{is} $"Background"$)}{
            $inpaint\_mask[pixel] \gets WHITEPIXEL$\;
        }{
            $inpaint\_mask[pixel] \gets BLACKPIXEL$\;
        }
    }
    $inpaint\_mask \gets \Dilate(inpaint\_mask, cur, type)$\;
    \KwRet $\ObjSplit{img, inpaint\_mask, cur, type}$\;
}

\Fn{\Dilate{mask, target, type}}{
    $mask \gets cv2.dilate(mask)$\;
    \If{$type \neq "Background"$}{
        \For{$pixel$ \textbf{in} $mask$}{
            \If{$pixel \in target$}{
                $mask[pixel] \gets BLACKPIXEL$\;
            }
        }
    }
    \KwRet $mask$\;
}

\Fn{\ObjSplit{img, inpaint\_mask, cur, type}}{
    $inpaint\_img \gets Lama(img, inpaint\_mask)$\;
    $inpaint\_map \gets Segmentation(inpaint\_img)$\;
    \eIf{$type \neq "Background"$}{
        $single \gets inpaint\_map[cur]$\;
    }{
        $single \gets inpaint\_map$\;
    }
        \KwRet $single$\;
}

\end{algorithm}

%% file: Chapters/algo-editor.tex
\SetKwFunction{RandMod}{RandMod}
\SetKwFunction{Editor}{Editor}
\SetKwProg{Fn}{Function}{:}{}

\begin{algorithm}[t]
\scriptsize
\caption{Layout Editor}\label{algo:editor}
\KwData{The mask of background $BG$, the masks set of single target objects $singles$ and the edited step budget $STEP$}
\KwResult{$semantic\_map$}

\Fn{\Editor{BG, singles, STEP}}{
    $index \gets 0$\;
    \While{$index < STEP$}{
        Select one of the $singles$ to apply any MR to it;
        $index \gets index + 1$;
    }
    $new\_layout \gets singles$ are stacked in order onto $BG$\;
    \KwRet $new\_layout$\;
}

\end{algorithm}

%% file: Chapters/algo-parser.tex
\SetKwFunction{ObjsExtract}{ObjsExtract}
\SetKwFunction{PosTag}{PosTag}
\SetKwProg{Fn}{Function}{:}{}

\begin{algorithm}[t]
\scriptsize
\caption{Caption Parser}\label{algo:parser}
\KwData{Image Description $ic$, Candidates $candis$ and Source Of Description $source$}
\KwResult{Target detection object $targets$ or $ic\_objs$}

\Fn{\ObjsExtract{ic, candis, source}}{
    $targets, ic\_objs \gets List(), \PosTag{ic}$\;
    \eIf{$source$ \textbf{is "GT"}}{
        \For{$cur$ \textbf{in} $ ic\_objs$}{
            \tcp{classifier is a classification model}
            $category \gets classifier(cur["name"])$\;
            \If{$category \in candis$}{
                $info \gets Dict()$\;
                $info["name"] \gets category$\;
                $info["num"] \gets candis["category"]["num"]$\;
                $targets.append(info)$\;
            }
        }
        \KwRet $targets$\;
    }{
        \For{$cur$ \textbf{in} $ ic\_objs$}{
            $cur["name"] \gets classifier(cur["name"])$\;
        }
        \KwRet $ic\_objs$\;
    }
}
\Fn{\PosTag{ic}}{
    \tcp{tagger is a part-of-speech tagging model}
    $tags, nouns, info, index \gets tagger(ic), List(), Dict(), 0$\;
    \While{$index < len(tags)$}{
        \If{($tags[i]["pos"]$ \textbf{is "ADJ" and } $index + 1 < len(tags)$ \textbf{and} $tags[i]["pos"]$ \textbf{is "NN"}) \textbf{or} $tags[i]["pos"]$ \textbf{is "NN"}}{
            $start \gets index$\;
            $end \gets start$\;
            \While{$end < len(tags)$ \textbf{and} $tags[end]["pos"] \textbf{is "NN"}$}{
                $end \gets end + 1$\;
            }
            $nnpm \gets ic[start: end]$\;
            $info["name"] \gets nnpm$\;
            \eIf{$index > 0$ \textbf{and the word before} $nnpm$ \textbf{is a number or} $\in$ ["a", "an"]}{
                $info["num"] \gets word2num(word)$\;
                $info["hasNum"] \gets True$\;
            }{
                $info["num"] \gets 1$\;
                $info["hasNum"] \gets False$\;
            }
            $nouns.append(info)$\;
            $index \gets end - 1$\;
        }
        $index \gets index + 1$\;
    }
    \KwRet $nouns$\;
}
\end{algorithm}

%% file: Chapters/evaluation.tex
\vspace{-2mm}
\section{Evaluation} 
\label{sec:eval}

We have developed \tool{} comprising 2043 Lines of Code (LoCs) in Python. Our evaluation of \tool{} is aimed at addressing the following research questions:

\begin{itemize}[itemsep=0pt,parsep=0pt,topsep=0pt,partopsep=0pt,left=0pt]
    \item \textbf{RQ1 (Reality\&Diversity):} What is the level of realism and diversity in the images generated by \tool?

    \item \textbf{RQ2 (Effectiveness):} How effective is \tool{} in identifying erroneous issues compared with other baselines?

    \item \textbf{RQ3 (Ablation Study):} What is the effectiveness of each MR in \tool? 
    
    \item \textbf{RQ4 (Error Categorization):} What types of description errors can \tool{} uncover?
\end{itemize}

\noindent \textbf{Experiment Setting.} All experiments are performed on a workstation with Ubuntu 22.04.3 LTS and 8 A100 GPU with 80GB memory. Comprehensive results and meticulous implementation details are accessible on our dedicated website~\cite{ourtool}. We generate 10 independent layout reconstructions per seed image. Subsequently, each refined semantic map undergoes a procedure to synthesize five new images. In the mask-to-image translation phase, we calibrate the classifier-free guidance strength to 1.3 and set the diffusion process to iterate for a total of 250 steps.

\noindent \textbf{Baselines.} We compare \tool to \metaic and \rome, which are the current SOTA metamorphic testing framework for testing IC systems. \metaic  established an independent object pool, generating descendant images with varying degrees of overlap by inserting random objects into ancestor images. It reports suspicious issues by comparing caption pairs between ancestor and descendant images. It includes two different patterns: non-overlapping(\metaic-NO) and partial-overlap(\metaic-PO). Following the settings in \rome, we choose a 30\% overlap as the default value for \metaic-PO. \rome also constructs ancestor-descendant image pairs. Differently, it generates descendant images by melting objects in the ancestor images. We follow the default setting of \rome.

\noindent \textbf{Dataset.} Consistent with the settings of previous works~\cite{rome, metaic}, we have selected the MSCOCO dataset, noted for its comprehensive collection of images and extensive annotations. Adhering to the experimental configurations of these previous works~\cite{rome, metaic}, we randomly selected 200 images and their corresponding annotations from MSCOCO to form our primary dataset. These images served as seeds for our test case generation. Using this default setting, \rome{}, \metaic-NO{}, \metaic-PO{}, and \tool{} produced a total of 362, 2,906, 2,906, and 10,000 test cases, respectively.

\noindent \textbf{IC Systems Under Test.} The subjects of our evaluation are seven prominent IC systems. Among these, one is the commercially available Azure AI Vision API~\cite{azure}, hereafter referred to as Azure. The remainder are open-source image-to-text models, specifically GIT~\cite{wang2022git}, BLIP~\cite{li2022blip}, BLIP2~\cite{li2023blip2}, ViT-GPT2~\cite{kumar2022illustrated}, OFA~\cite{wang2022ofa}, and VinVL~\cite{zhang2021vinvl}. For each of these, we have employed the pre-trained models as released by the original authors for our testing purposes.

\vspace{-2mm}
\subsection{RQ1: What is the level of realism and diversity in the images generated by \tool?}

\begin{figure*}[t]
    \centering
    \begin{subfigure}{0.315\textwidth}
        \centering
        \includegraphics[width=1\textwidth]{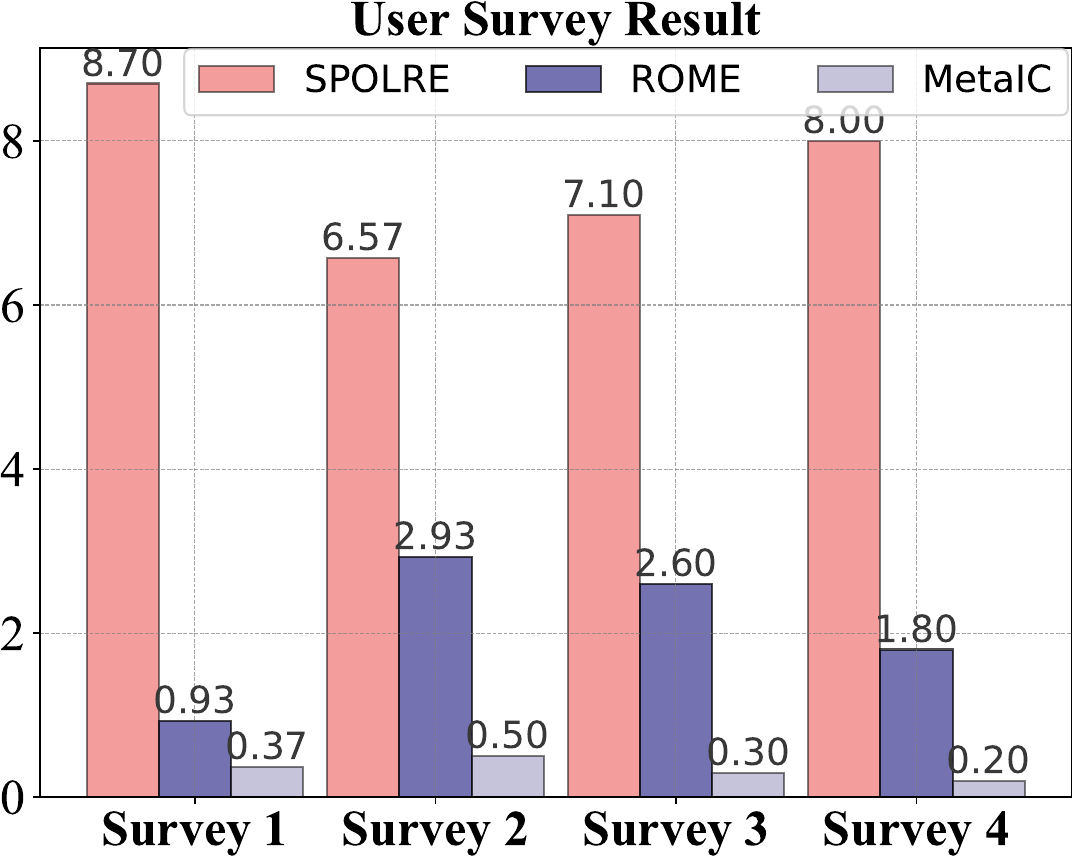}
        \caption{The average number of people votes for images from \tool and baselines.}
        \label{fig:study}
    \end{subfigure}
    \hspace{2mm}
    \centering
    \begin{subfigure}{0.32\textwidth}
        \centering
        \includegraphics[width=1\textwidth]{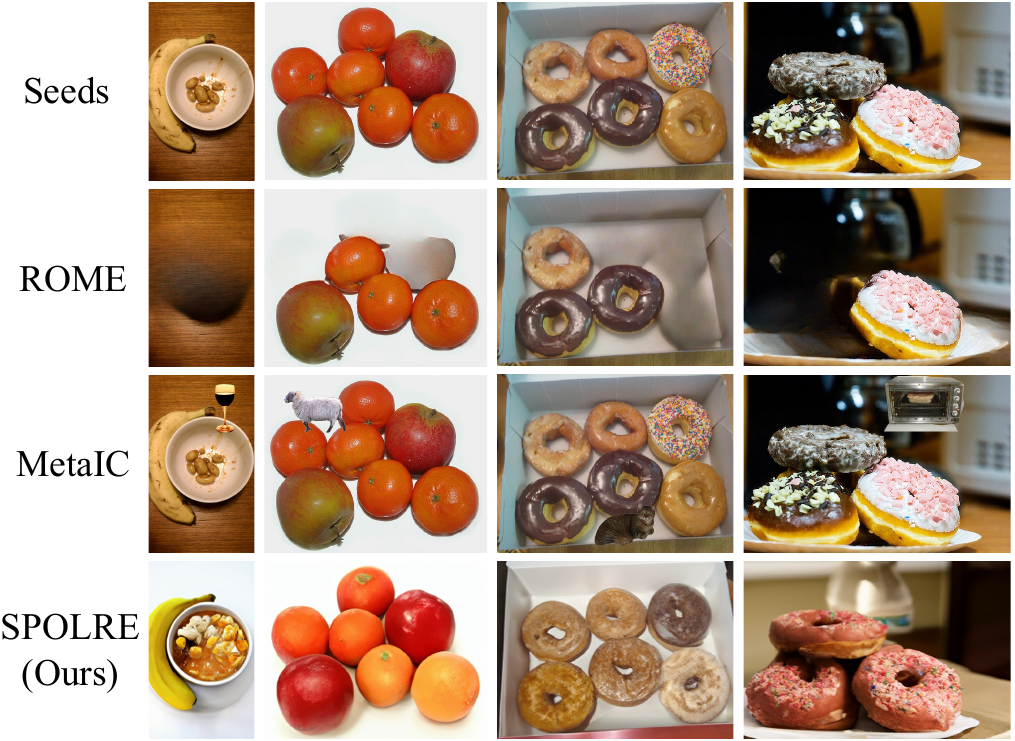}
        \caption{Comparison of part results of \tool and baseline under the same seeds.}
        \label{fig:compare}
    \end{subfigure}
    \hspace{2mm}
    \begin{subfigure}{0.25\textwidth}
        \centering
        \includegraphics[width=1\textwidth]{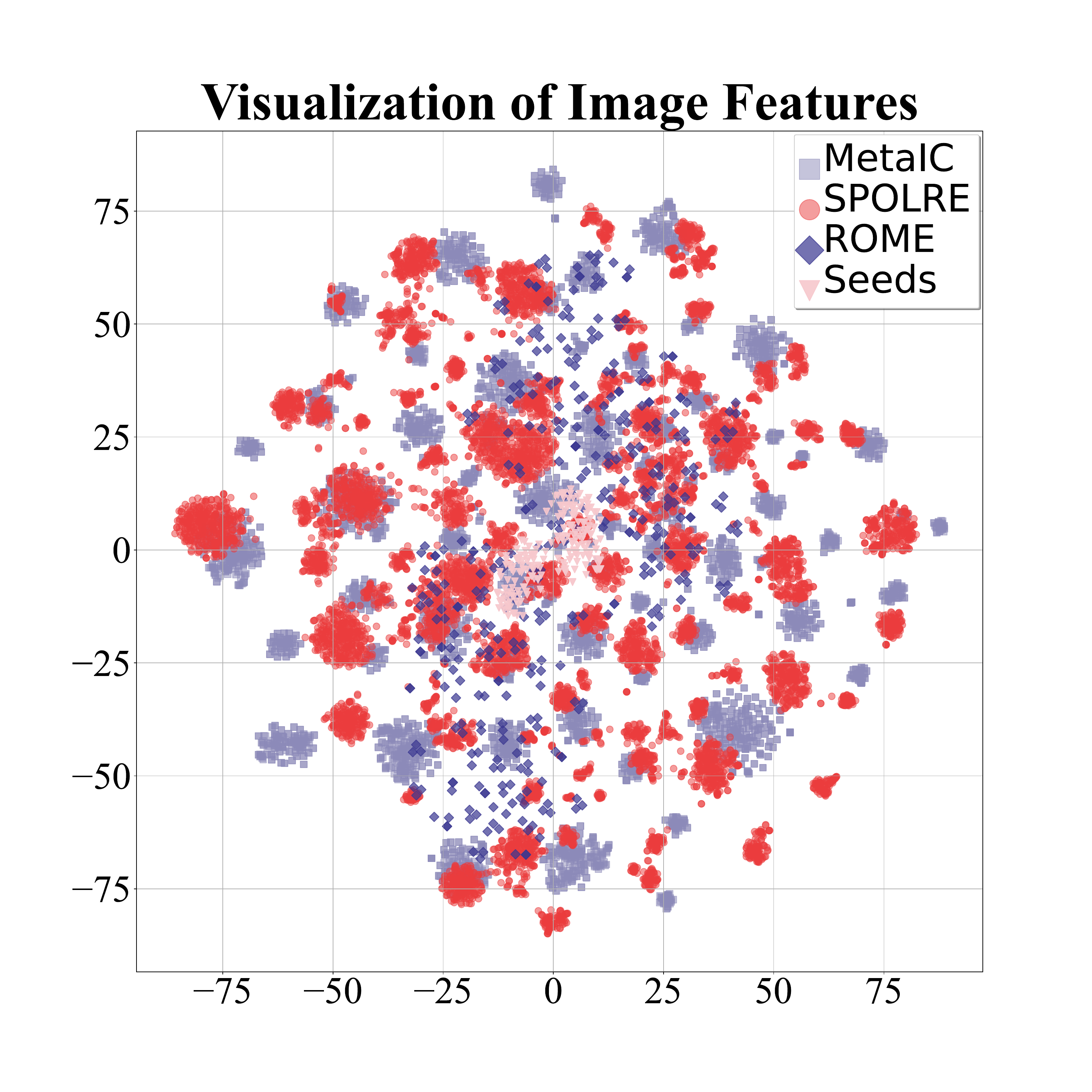}
        \caption{The distribution of generated images in the feature space.}
        \label{fig:scatter}
    \end{subfigure}
    \caption{Research results on the authenticity and diversity of generated images.} 
    \label{fig:rq1}
\end{figure*}

To evaluate the realism of images crafted by \tool{} in comparison to state-of-the-art methods such as \rome{} and \metaic{}, we conducted a comprehensive user study with 120 volunteer participants. The study was structured into four distinct surveys, each designed to evaluate the impact of different metamorphic transformations on user perception. By focusing each survey on a specific transformation, we aimed to isolate the effects and gather precise data. Outputs generated by the three methods were assessed using 10 randomly selected seeds, resulting in a comparative analysis of 30 images per method. Each survey asked a consistent set of questions to ensure comparability, allowing us to directly assess the impact on user responses. Conducted with different groups of users to avoid bias and ensure diverse perspectives, the surveys were not conducted simultaneously to prevent overlap in participant experiences. Participants identified the more realistic and natural image from each pair and rated their confidence on a scale from one (very uncertain) to five (very certain). This  framework was chosen to enhance the study's validity. We gathered 120 valid responses, 30 for each survey, highlighting the comparative effectiveness of \tool{} against SOTA methods. Participants were recruited through university mailing lists to ensure diversity. While our primary focus was on transformations, we acknowledge the importance of evaluating the realism of the original images as a baseline and will include this in future work. A sample of the responses can be explored at ~\cite{userstudy}, and \autoref{fig:compare} illustrates the comparative results of partially generated images alongside the seeds, further validating our approach.

As depicted in \autoref{fig:study}, our survey results reveal that over 75\% of participants perceive images produced by \tool as more natural than those by baselines in each survey iteration. To objectively assess this perceived difference in realism, we conducted a t-test~\cite{kim2015t} comparing images from \tool and baselines. The analysis produced a \(t\)-value of 4.05 and a \(p\)-value of 0.0001, emphatically rejecting the null hypothesis at a 0.05 significance level in a two-tailed hypothesis test. This finding indicates a statistically significant disparity in naturalness between the image sets. The positive and substantially high \(t\)-value suggests that \tool's images are more naturalistic than those from baselines. Consequently, this endorses the utility of \tool-generated images in evaluating IC systems' ability to describe images accurately. 

Additionally, we embed the generated images into a feature space using ResNet50~\cite{he2016deep} and visualize the embedding results with t-SNE~\cite{van2008visualizing}. As shown in \autoref{fig:scatter}, under the condition of keeping the objects within the images unchanged, the coverage of images generated by \tool is comparable or even superior to the two baseline methods, whereas the latter two modify the original semantics of the seed images through operations of inserting or deleting objects.
Overall, the user study outcomes affirm the validity and practicality of \tool in the context of image realism and diversity.

\subsection{RQ2: How effective is \tool in detecting error descriptions compared with other baselines?}

In this RQ, we focus on evaluating the efficacy of \tool in identifying erroneous descriptions by IC systems. Adhering to methodologies established in previous research~\cite{rome}, we randomly selected 200 images from MSCOCO to serve as inputs for generating test cases using various approaches. Each approach then employed IC systems to create captions for their test cases, with error detection based on their respective criteria. For clarity in our explanation, we adopt the terminologies used in \metaic and \rome: We designate the original input image as the ancestor ($I_{a}$) and the corresponding generated image as the descendant ($I_{d}$). The caption pair related to the image pair $<I_{a}, I_{d}>$ is represented as $<D_{a}, D_{d}>$.

To evaluate the effectiveness of error detection, we measure the \textbf{Precision} of reported errors. An error detection is considered positive if the tools report any discrepancies. Mirroring the methodology from \cite{rome}, two independently reviewed reported errors to identify True Positive (TP) and False Positive (FP) instances. To verify the reliability of this manual labeling process, we computed the Cohen's Kappa~\cite{cohen1960coefficient} coefficient, which yielded a value of 0.83. This score signifies a high level of agreement between the reviewers, reinforcing the consistency and accuracy of their assessments.

\begin{table}[t]
\caption{Precision of \tool and Baselines}
\vspace{-3mm}
\label{table:precision}
\begin{adjustbox}{width=\linewidth}
\begin{tabular}{c|cccc}
\toprule
\textbf{IC Systems}           & \multicolumn{4}{c}{\textbf{Testing Methods}}                                \\ 
\midrule
\textbf{} & \textbf{ROME} & \textbf{MetaIC-NO} & \textbf{MetaIC-PO} & \textbf{\tool}      \\
\textbf{Azure}      & 64.56\%(51/79)   & 83.87\%(1992/2375)    & 80.93\%(1867/2307)    & \textbf{95.15\%}(6236/6554)   \\
\textbf{GIT}        & 61.96\%(57/92)   & 89.07\%(2331/2617)    & 87.16\%(2247/2578)    & \textbf{92.75\%}(4913/5297)   \\
\textbf{BLIP}       & 71.72\%(71/99)   & \textbf{83.46\%}(2033/2436)    & 81.37\%(1966/2416)    & 80.93\%(2957/3654)   \\
\textbf{BLIP2}      & 65.59\%(61/93)   & 75.67\%(1608/2125)    & 72.17\%(1499/2077)    & \textbf{89.16\%}(3505/3931)   \\
\textbf{Vit-GPT2}   & 65.60\%(82/125)  & \textbf{95.70\%}(2715/2837)    & 94.28\%(2669/2831)    & 94.28\%(5174/5488)   \\
\textbf{OFA}        & 64.86\%(48/74)   & 84.49\%(1988/2353)    & 80.94\%(1890/2335)    & \textbf{94.80\%}(3934/4150)   \\
\textbf{VinVL}      & 70.65\%(65/92)   & 83.99\%(2030/2417)    & 82.97\%(1997/2407)    & \textbf{90.09\%}(4825/5356)    \\ 
\hline
\textbf{Average}    & 66.51\%(435/654) & 85.65\%(14697/17160)   & 83.39\%(14135/16951)  & \textbf{91.62\%}(31544/34430) \\ 
\bottomrule
\end{tabular}
\end{adjustbox}
\vspace{-3mm}
\end{table}

Results are detailed in \autoref{table:precision}. Our analysis reveals that \tool's precision outperforms the baseline metrics across five systems, averaging around 91.62\%. Intriguingly, when utilizing the same seed images, \tool identified a total of 31,544 errors post-manual review. In contrast, \rome, as well as both variants of \metaic (non-overlapping and partial-overlap), reported considerably fewer errors—654, 17,160, and 16,951, respectively. This significant discrepancy stems from \tool's unique approach of redrawing the seed image rather than directly adding or removing objects, thereby maintaining the original image's semantic integrity. Additionally, \tool employs a controlled diffusion model for mask-to-image transformation. This method not only retains the objects from the ground truth but also alters aspects like object texture. Therefore, \tool is capable of generating diverse test scenarios, significantly enhancing its potential to uncover a broader spectrum of issues.

The efficiency of the two error detection methods employed by \metaic pales in comparison to that of \tool. Specifically, the two modes of \metaic only identified 14,697 and 14,135 errors, respectively, after manual review. This relative inefficiency is attributed to their approach in generating test cases. \metaic generates descendant images by directly inserting extraneous objects into the images, which often results in the creation of visually unnatural samples. Consequently, IC systems frequently fail to adequately describe these artificially inserted objects, leading to missed error detections.

\begin{figure}[t]
    \centering
    \begin{subfigure}{0.325\linewidth}
        \centering
        \includegraphics[width=\textwidth]{Figures/rome-azure-a.png}
        \caption{}
        \label{fig:raa}
    \end{subfigure}
    \begin{subfigure}{0.325\linewidth}
        \centering
        \includegraphics[width=\textwidth]{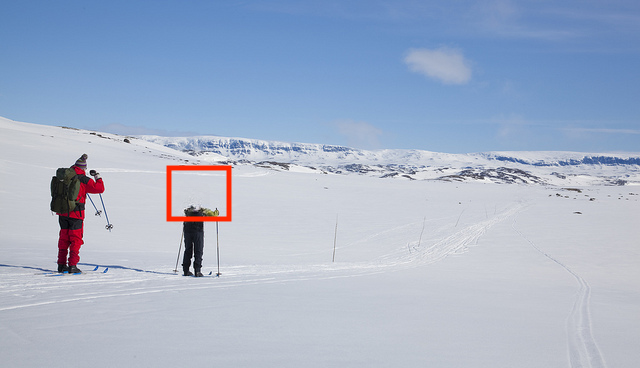}
        \caption{}
        \label{fig:rab}
    \end{subfigure}
    \begin{subfigure}{0.325\linewidth}
        \centering
        \includegraphics[width=\textwidth]{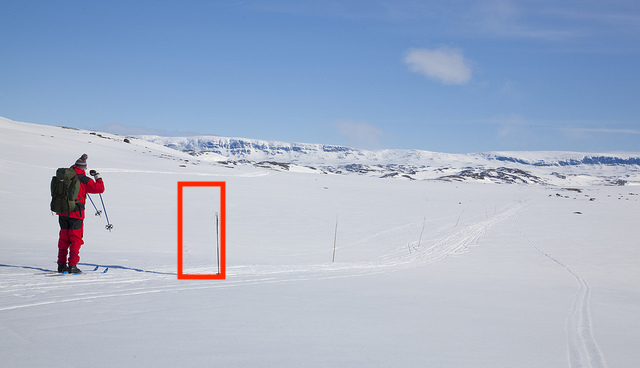}
        \caption{}
        \label{fig:rac}
    \end{subfigure}

    \caption{An example of a problem with \rome. $D_a$\&$D_b$: A person skiing on a snowy mountain. $D_c$:  A snow covered landscape with a person on a ski slope. \rome reported: $<D_a, D_b>$: Normal, $<D_b, D_c>$: Abnormal}
    \label{fig:rome-issue}
    \vspace{-2mm}
\end{figure}

Additionally, \rome exhibits lower precision, which we attribute to the random selection of seed images. Using the Lama technique for larger object deletion can result in issues like incomplete removal or unnatural images due to inadequate context as shown in \autoref{fig:motivation}, complicating error detection.
Furthermore, \rome's selection of objects for deletion is based on MSCOCO annotations, which may not always coincide with the image's focal point. This misalignment is illustrated in \autoref{fig:rome-issue}: In the pair $<I_a, I_b>$, a backpack is removed, but since it is not the central element, captions $D_a$ and $D_b$ remain identical, leading to no detected error. Conversely, in $<I_b, I_c>$, the removal of a person is significant, and the discrepancy between $<D_b, D_c>$ results in an error report. Moreover, \rome detects the fewest number of errors due to its generation logic. The number of image pairs \rome can produce is contingent on the object count in the seed images. For a seed with $n$ detectable objects, \rome theoretically generates $3^n - 2^n$ pairs per seed, leading to a constrained total number of test cases. In particular, we note that in order to reduce the problem of incomplete deletion, \rome always retains the largest object in an image. This further reduces the number of test cases that ROME can generate. This selective approach likely contributes to the lower number and precision of errors detected by \rome.


\subsection{RQ3: What is the effectiveness of each MR in \tool?}

In this RQ, our aim is to assess the effectiveness of the MRs established in \tool{}. To achieve this, we conducted an ablation study with four distinct variants of \tool, designated as \tool-MR$i (i = 1, 2, 3, 4)$. Each variant exclusively implements one of the MRs. For the experiment's design, identical seed images were input into each variant, and five layout reconstructions were performed for each $I_{a}$. Each reconstruction involved a single modification step consistent with the specific MR of that variant. The resulting images, denoted as $I_{m-n}$, were generated based on these semantic maps, where \(m\) represents the MR index, and \(n\) signifies the generated image index. Images sharing the same \(m\)-value but varying in \(n\)-value indicate their origin from the same variant applied to the common $I_{a}$, albeit with differing target objects and transformation extents under the given MR. Subsequently, 1000 $<I_{a}, I_{m-n}>$ pairs were produced for each variant. We then collected captions for these image pairs and executed the error detection process.

The results for each MR are detailed in \autoref{table:ablation}. In total, we identified 9,404 anomalies across the four variants, with their respective shares of abnormalities being approximately 24.64\%, 25.10\%, 24.24\%, and 26.02\%. This pattern is also mirrored in the performance of individual IC systems, suggesting that each defined MR is vital in detecting issues, and none of them can be deemed redundant or superfluous.

Additionally, in the ablation experiment encompassing 28,000 tests across seven IC systems, we observed an error rate of about 33.59\%. This contrasts with the 45.06\% error rate encountered in our prior experiment, which involved 70,000 tests. This disparity implies that integrating multiple MRs in \tool{} is more effective than relying on a single MR, underscoring the synergy achieved through their combination.

\begin{table}[t]
\caption{The Results of Ablation Experiment}\vspace{-2mm}
\label{table:ablation}
\begin{adjustbox}{width=0.95\linewidth}
\begin{tabular}{c|cccc}
\toprule
\textbf{IC Systems} & \multicolumn{4}{c}{\textbf{Variants of \tool}}                            \\ 
\midrule
\textbf{}           & \textbf{\tool-MR1} & \textbf{\tool-MR2} & \textbf{\tool-MR3} & \textbf{\tool-MR4} \\
\textbf{Azure}      & 443                  & 451               & 453              & 456             \\
\textbf{GIT}        & 371                  & 339               & 352              & 379             \\
\textbf{BLIP}       & 247                  & 239               & 223              & 259             \\
\textbf{BLIP2}      & 289                  & 278               & 275              & 303             \\
\textbf{Vit-GPT2}   & 332                  & 383               & 357              & 409             \\
\textbf{OFA}        & 263                  & 292               & 262              & 289             \\
\textbf{VinVL}      & 372                  & 378               & 358              & 352             \\ \hline
\textbf{Total}      & 2317                 & 2360              & 2280             & 2447            \\ 
\bottomrule
\end{tabular}
\end{adjustbox}
\vspace{-3mm}
\end{table}

\subsection{RQ4: What types of description errors can our tools find?}

\begin{figure*}[t]
    \centering
    \includegraphics[width=0.95\textwidth]{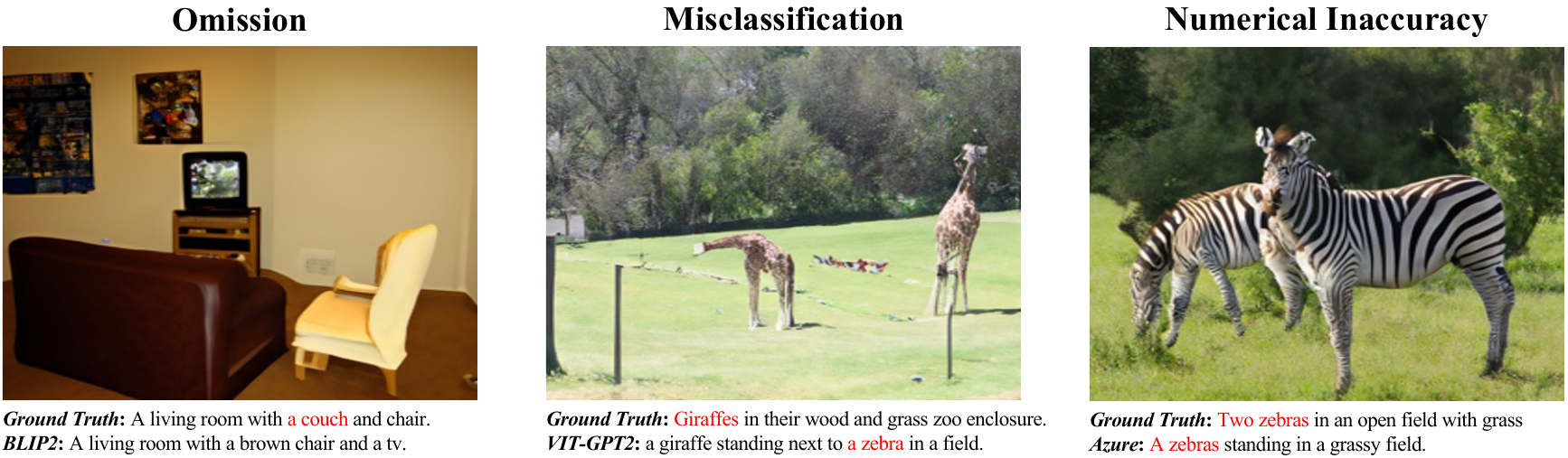}
    \vspace{-2mm}
    \caption{The cases of real errors found by \tool.}
    \vspace{-4mm}
    \label{fig:case}
\end{figure*}

In this RQ, we delve into the specific types of errors identified by \tool. Upon a detailed examination of the error reports, we categorize the errors into three distinct types: \textbf{Omission}, \textbf{Misclassification}, and \textbf{Numerical Inaccuracy}. Illustrative examples of each error type detected by \tool are presented in \autoref{fig:case}. It is important to note that a single caption pair $<D_{a}, D_{d}>$ may encompass multiple error types.

\begin{figure}[t]
    \centering
    \includegraphics[width=0.925\linewidth]{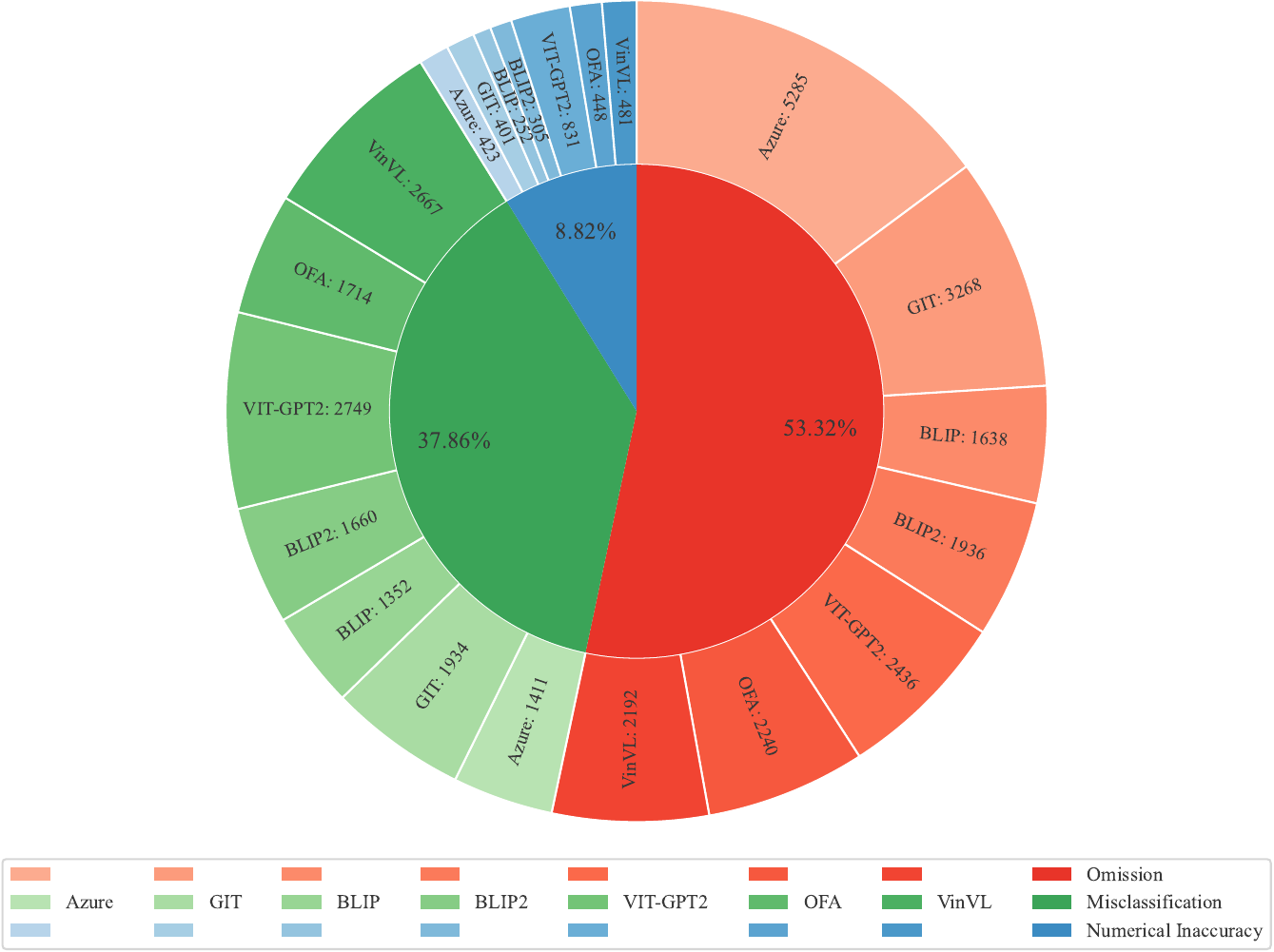}
    \vspace{-2mm}
    \caption{The proportion of three types of errors and the number reported in each IC system.}
    \vspace{-4mm}
    \label{fig:chart}
\end{figure}

In an analysis covering seven IC systems, \tool identified a total of 31,544 manually reviewed exceptions, uncovering 35,623 distinct errors. The distribution of errors across the categories of Omission, Misclassification, and Numerical Inaccuracy is 53.32\%, 37.86\%, and 8.82\%, respectively. A detailed breakdown of these findings is depicted in \autoref{fig:chart}.

We can observe that the incidence of Numerical Inaccuracy errors is markedly lower compared to the other two categories, Omission and Misclassification. This discrepancy can be attributed to two primary factors. Firstly, the detection process for Numerical Inaccuracy is conditional; it is initiated only when the IC system explicitly specifies the number of objects in an image. Consequently, the opportunity for detecting numerical inaccuracies is inherently less frequent than for the other error types. Secondly, the design ethos of \tool primarily focuses on modifying the spatial arrangement and attributes of objects within an image, as well as altering the background context, rather than changing the quantity of the objects. This approach implies that detecting Numerical Inaccuracy is not one of \tool's forte, as it does not typically induce variations in the number of objects, which is a key aspect in identifying such errors.

Additionally, as shown in \autoref{fig:chart}, we find a notably higher frequency of Omission errors (such as the example in \autoref{fig:case}) in Azure compared to other systems. As discussed in \rome~\cite{rome}, Azure generally refrains from describing objects that appear unnatural. Given that the transformations applied by \tool{}, such as translations, rotations, and scaling, are executed randomly, even though we confine the magnitude of each modification, these operations can still lead to atypical object layouts. This aspect of our methodology is likely the reason behind the disproportionately high count of Omission errors observed in Azure.

%% file: Chapters/discussion.tex
\section{Threats To Validity}

\subsection{Impact of Incorrect Ground Truths}
Ground truth misclassifications may lead to False Positives in our error detection process. We attempt to mitigate this by filtering anomalies based on semantic segmentation, yet this might overlook some errors. Future work will explore ground truth-free optimization to improve this aspect.

\subsection{Uncertainty in User Surveys}
To ensure the reliability of our user survey assessing image naturalness, all users received clear instructions before participating, and we included such information in the questionnaires as well. Furthermore, participants were required to spend a minimum of 30 seconds per question and compare images with their originals. These measures aimed to reduce random responses and enhance the survey's credibility.

%% file: Chapters/related-work.tex
\section{Related Work}
\label{sec:relwork}

\subsection{Robustness of AIGC}

The proliferation of Artificial Intelligence Generated Content (AIGC) powered by advancements in large multi-modal models has made the robustness of these technologies a critical area of inquiry. As these models become integral to diverse applications, ensuring their reliability and error resilience is paramount.

In image generation, thorough investigations have been conducted into the robustness of large multi-modal models. Studies like ~\cite{bird2023typology} have explored the typological errors and distortions that can occur due to biased or incomplete training data within text-to-image frameworks. Furthermore, ~\cite{carlini2023extracting} has highlighted crucial concerns about data integrity in diffusion models, pointing to the need for robust methods that preserve data fidelity and prevent corruption. The potential for embedding unintentional backdoor vulnerabilities into these models, which could be exploited under specific conditions, has been examined in depth in works such as ~\cite{chou2023backdoor, huang2023zero, chou2023villandiffusion}, emphasizing the importance of thorough robustness testing.

Parallel concerns arise in the robustness of text generation. The phenomenon of hallucinations, where large multi-modal models generate content that is plausible yet factually incorrect, has been a significant focus. The conditions facilitating these hallucinations have been scrutinized in studies like ~\cite{li2022pre, dziri2022origin, zheng2023does, rawte2023troubling}. Efforts to mitigate these errors and enhance the factual accuracy of generated text are documented in ~\cite{lee2023factuality, penedo2023refinedweb, zhou2023lima, elaraby2023halo, li2023toolaugmented}.

Innovative methods for probing the limits of these models' robustness have also been introduced, such as those by ~\cite{deng2023jailbreaker}, which develop prompts designed to test and expose the robustness of controlled conditions within these models.

Our study extends this discussion to the robustness of large multi-modal models in a multimodal context, integrating both textual and visual outputs. We build upon foundational studies like ~\cite{metaic} and ~\cite{rome}, focusing on how these large systems manage the complex interplay between different types of data inputs and outputs. This exploration not only deepens our understanding of specific risks but also fosters the development of more sophisticated and robust multi-modal systems capable of delivering reliable, error-resilient content in a variety of practical applications.

\subsection{Metamorphic Testing}

Metamorphic Testing (MT), as described in ~\cite{chen2020metamorphic}, addresses the critical challenges associated with the test oracle problem, particularly in scenarios where outcomes are inherently ambiguous. This testing methodology is crucial in situations where it is not feasible to predetermine correct outputs for given inputs, making it an indispensable tool in the evaluation of complex software systems.

The applications of MT are diverse and have been explored across various domains to enhance reliability and robustness. For instance, ~\cite{xiao2022metamorphic} focused on the application of MT in the context of deep learning compilers. These compilers, which are essential for optimizing machine learning operations, often involve intricate transformations that can introduce subtle errors, undetectable by traditional testing methods. MT provides a framework to reveal these errors by testing how changes in input affect the output in predictable ways.

Another significant application of MT has been in the domain of textual content moderation software, as explored by ~\cite{wang2023mttm}. In this context, MT helps in identifying biases and errors in the moderation algorithms by applying transformations to the input texts and analyzing the consistency of the outputs. This method ensures that content moderation systems behave as expected across a variety of inputs, including those crafted to test the limits of the system's robustness. Additionally, ~\cite{li2024halluvaultnovellogicprogrammingaided, wang2024metmap} propose a metamorphic testing approach to detect issues such as fact-conflicting hallucinations and vector matching errors in large language models.

Moreover, MT has been extensively applied in multimodal image-text tasks, which involve interactions between textual and visual data. Studies such as ~\cite{wang2020metamorphic, metaic, rome} have utilized MT to test IC systems. These systems, which generate textual descriptions for images, can benefit from MT by verifying that changes in image attributes lead to appropriate changes in the generated captions, thereby testing the system's ability to handle diverse scenarios.

Our contribution to this field lies in enhancing the naturalness of test cases and improving issue detection in IC systems using MT. By strategically selecting fewer seeds for generating test cases, we have developed a methodology that not only reduces the computational overhead but also improves the efficiency of detecting significant discrepancies in system outputs. We focus on creating more natural and realistic transformations, ensuring that the test cases are not only effective but also reflective of real-world scenarios. This approach helps in identifying more subtle, yet critical issues that might be overlooked by more straightforward test cases.

In conclusion, MT continues to be a powerful method for validating the robustness and reliability of software systems, especially those involving complex interactions between different types of data and those operating under conditions where traditional testing methods fall short. Our enhancements in test case design and implementation further the capabilities of MT, making it a more effective tool in the ongoing effort to improve the accuracy and fairness of automated systems.

\subsection{Image Generation by Diffusion Models}

Diffusion models have become a cornerstone in image generation, significantly advancing the synthesis of detailed images from textual descriptions. These models combine the generative prowess of advanced language models with diffusion processes, enhancing their ability to interpret and visualize complex textual inputs effectively.

Recent studies, such as ~\cite{ramesh2022hierarchical, betker2023improving, chen2023textdiffuser, chen2023textdiffuser2}, have focused on refining these models to better decode abstract concepts into visually captivating images. The integration of artistic styles, explored in ~\cite{zhang2023inversion, hamazaspyan2023diffusion}, allows diffusion models to produce images in a variety of artistic expressions, achieving a balance between style and content coherence.

Moreover, research efforts like ~\cite{rombach2022high, saharia2022image} are dedicated to improving image resolution and quality, ensuring high-definition outputs suitable for demanding applications. In the area of image restoration, studies such as ~\cite{rombach2022high, li2022sdm, lugmayr2022repaint} highlight the models’ ability to seamlessly repair or reconstruct damaged images, which is crucial for archival restoration and media recovery.

Diffusion models are not only enhancing image generation quality but are also expanding into applications that require dynamic image responses, such as in virtual reality. Their development continues to push the boundaries of how AI can integrate into and enhance digital media and creative industries.


%% file: Chapters/conclusion.tex
\section{Conclusion}
\label{sec:conclusion}

This paper presents \tool, an innovative semantic preserving object layout reconstructor designed for IC systems, developed to overcome the shortcomings of current test case generation methods. Our strategy integrates image segmentation, editing, and image-to-image translation. By reconstructing semantic maps of seed images, \tool produces a wider array of test scenarios. Our approach, distinct from traditional methods of object addition or deletion on static backgrounds, utilizes a diffusion model to redraw object environments, resulting in more vibrant and diverse images. Tested across seven IC systems, \tool demonstrates superior performance in both image generation quality and diversity. \tool also excels in identifying more error descriptions using the same seeds. Future work will explore additional 
object-level 
modifications, methods independent of ground truth reliance, and further testing on diverse IC systems to enhance their accuracy.